*Review*

# Electroluminescence and electron avalanching in two-phase detectors

## A. Buzulutskov [1,2]

[1] Budker Institute of Nuclear Physics, 630090 Novosibirsk, Russia
[2] Novosibirsk State University, 630090 Novosibirsk, Russia
Correspondence: A.F.Buzulutskov@inp.nsk.su

**Abstract:** Electroluminescence and electron avalanching are the physical effects used in two-phase argon and xenon detectors for dark mater search and neutrino detection, to amplify the primary ionization signal directly in cryogenic noble-gas media. We review the concepts of such light and charge signal amplification, including a combination thereof, both in the gas and in the liquid phase. Puzzling aspects of the physics of electroluminescence and electron avalanching in two-phase detectors are explained and detection techniques based on these effects are described.

**Keywords:** two-phase detectors for dark matter search; liquid noble-gas detectors; proportional electroluminescence in noble gases; electron avalanching at cryogenic temperatures.

## 1. Introduction

The ultimate goal for liquid noble-gas detectors is the development of large-volume detectors of superior sensitivity for rare-event experiments. A typical deposited energy in such experiments might be rather low: of the order of 0.1 keV in coherent neutrino-nucleus scattering experiments, 1-100 keV in those of dark matter search and ≥100 keV in those of astrophysical and accelerator neutrino detection. Accordingly, the primary ionization and scintillation signals, produced by a particle in noble-gas liquid, have to be amplified in dense noble-gas media at cryogenic temperatures.

The idea of two-phase detectors [1] is based on the requirement of amplification of such weak signals (see reviews [2,3]). In particular, the scattered particle produces in the liquid phase two types of signals: that of primary scintillation recorded promptly (S1 signal) and that of primary ionization recorded with delay (S2 signal). While for readout of the S1 signal it is sufficient to use optical devices like PMTs and SiPMs having their own electronic amplification, to record the S2 signal one has to amplify it in the detection medium directly (in case of the low deposited energy). In two-phase detectors this is done in the following way: the primary ionization electrons, after drifting in the liquid phase, are emitted into the gas phase where their (S2) signal is amplified. In "classic" two-phase detectors, the way to amplify the S2 signal is provided by electroluminescence (secondary scintillation) produced by drifting electrons under high electric fields. The second possible way to amplify the S2 signal is that of electron avalanching (electron multiplication) at high electric fields. In the following, we refer to these two ways of amplification as those of light and charge signal amplification, respectively.

This review is dedicated to the issue of S2 signal amplification. We review all known concepts of light and charge signal amplification (i.e. those based on electroluminescence and electron avalanching in cryogenic noble-gas media), including a combination thereof, both in the gas and in the liquid phase. We explain puzzling aspects of the physics of electroluminescence and electron avalanching in two-phase detectors and describe detection techniques based on these effects.

## 2. Basic concepts of signal amplification in two-phase detectors

Many methods for light and charge signal amplification in two-phase detectors have been proposed, and some of these are described in reviews [2,3,4,5,6]. Among this variety, two basic concepts can be distinguished, that most other concepts come from.

The basic concept of light signal amplification is that of the "classic" two-phase detector (see [3]); it is provided by the effect of proportional electroluminescence (EL): see Figure 1. Here each drifting electron of the ionization (S2) signal emits light in the EL gap under high electric field, the light intensity being roughly proportional to the electric field. This concept splits into two sub-concepts: with indirect and direct optical readout. For the former, the optical readout goes via a wavelength shifter (WLS).

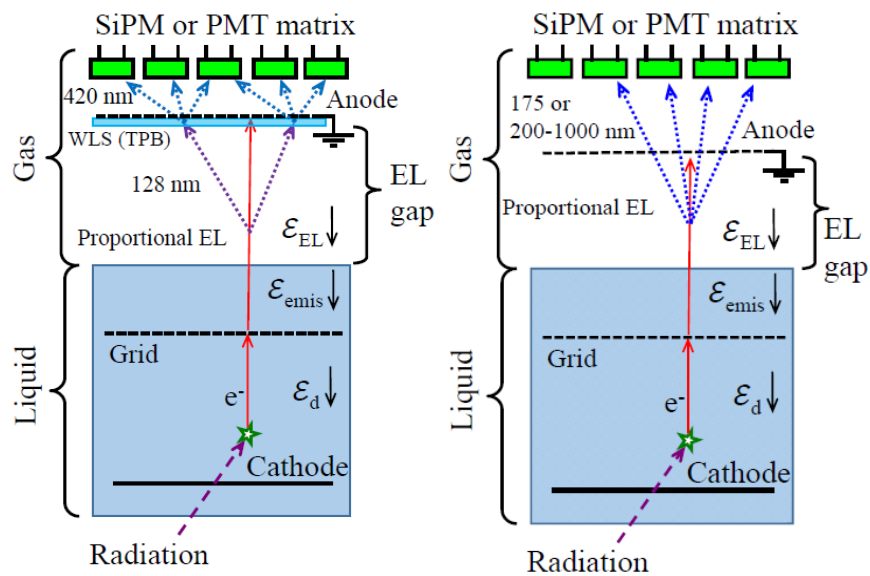

**Figure 1.** Basic concepts of light signal amplification in two-phase detectors, using proportional electroluminescence (EL) in the EL gap. Left: with indirect optical readout via wavelength shifter (WLS), in two-phase Ar using excimer emission in the VUV at 128 nm. Right: with direct optical readout, either in two-phase Xe using excimer emission in the VUV at 175 nm, or in two-phase Ar using neutral bremsstrahlung (NBrS) emission in the non-VUV range (at 200-1000 nm).

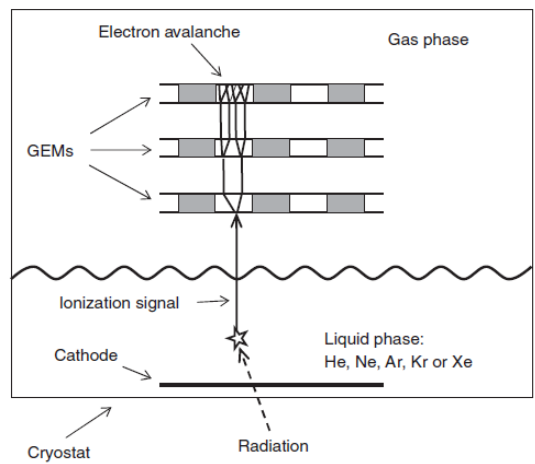

**Figure 2.** Basic concept of charge signal amplification in two-phase detectors, using GEM-like structures [4].

The basic concept of charge signal amplification in two-phase detectors is that using a GEM-like structure (Gas Electron Multiplier, GEM) in the gas phase that provides electron avalanching within the GEM holes at cryogenic temperatures: see Figure 2 and review [4].

In sections 3-6, we describe in detail the physics and state-of-the-art of these basic concepts, as well as consider other (most developed) concepts of signal amplification in cryogenic noble-gas media.

## 3. Light signal amplification in the gas phase of two-phase detector, using proportional electroluminescence

### 3.1. Three EL mechanisms

According to modern concepts, there are three EL mechanisms responsible for proportional EL in gaseous Ar and Xe: that of excimer ($Ar_2^*$ or $Xe_2^*$) emission in the VUV ("ordinary" EL), that of emission due to atomic transitions in the Near Infrared (NIR), and that of neutral bremsstrahlung (NBrS) emission in the UV, visible and NIR range. The energy level diagrams, appropriate reactions and their constants, relevant to the former two mechanisms, are presented in Figure 3 and Table 1. Photon emission spectra for all three mechanisms and their absolute EL yields are shown in Figures 4 and 5 respectively for Ar [7]. For other noble gases the emission spectra and EL yields look alike [4,5]. For Ar, the EL mechanisms are described below in detail.

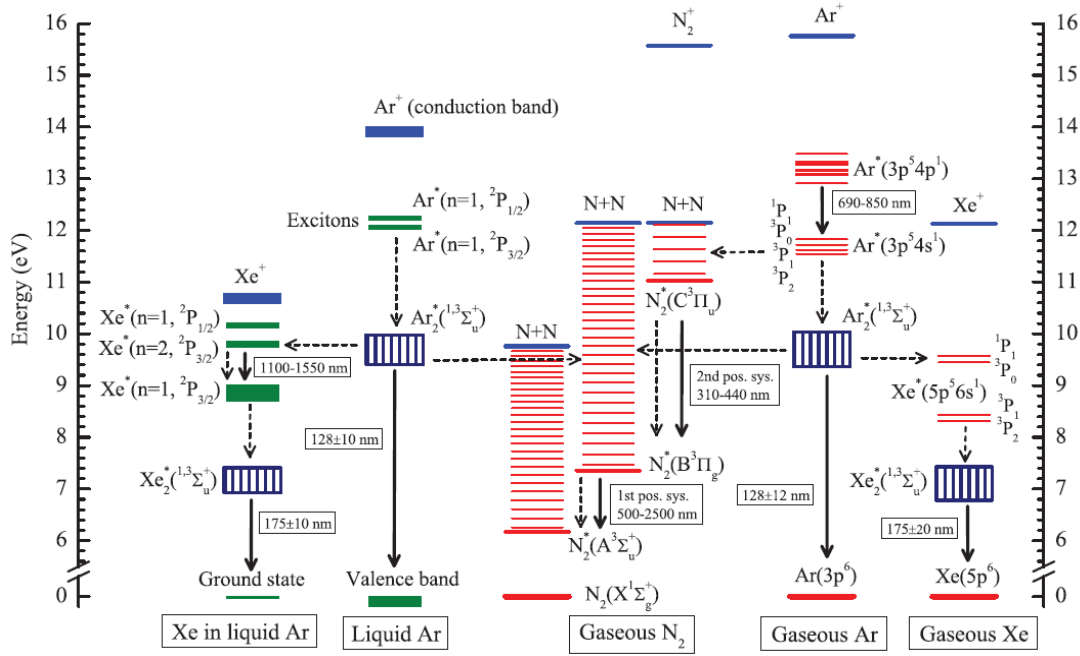

**Figure 3.** Energy levels of the lower excited and ionized states relevant to the ternary mixture of Ar doped with Xe and N₂ in the two-phase mode [7]. These are shown for gaseous Ar, gaseous N₂, gaseous Xe, liquid Ar and Xe in liquid Ar. The solid arrows indicate the radiative transitions observed in experiments and relevant to the present review (i.e. when the excitation is induced by ionization or electroluminescence). The numbers next to each arrow show the photon emission band of the transition, defined by major emission lines or by full width of the emission continuum. The dashed arrows indicate the most probable non-radiative transitions induced by atomic collisions for Ar, Xe and N₂ species and their pair combinations in the gas and liquid phase.

Proportional EL is characterized by a reduced EL yield ($Y/N$), which by definition is the number of photons produced per drifting electron per atomic density and per unit drift path. It is typically plotted as a function of the reduced electric field ($E/N$), the latter being expressed in Townsends: 1 Td = $10^{-17}$ V cm² atom⁻¹, corresponding to 0.87 kV/cm in gaseous Ar at 1.00 atm and 87.3 K.

**Table 1.** Basic reactions of excited species relevant to the performance in the two-phase mode, namely in Ar in the gas and liquid phase, doped with Xe (1000 ppm in the liquid and 40 ppb in the gas phase) and $N_2$ (50 ppm in the liquid and 135 ppm in the gas phase), their rate (k) or time ($\tau$ ) constants reported in the literature and their time constants reduced to given atomic densities at 87 K ($\tau_{TP}$ ), in particular for Ar to that of $8.63 \times 10^{19}$ cm$^{-3}$ and $2.11 \times 10^{22}$ cm$^{-3}$ in the gas and liquid phase, respectively [7]. The references presented in the table are those of [7].

| No. | Reaction | k or $\tau$ | T | Reference | $\tau_{TP}$ |
|---|---|---|---|---|---|
| | Gaseous Ar + Xe (40 ppb) + $N_2$ (135 ppm) | | | | |
| (1) | $Ar^*(3p^54s^1) + 2Ar \rightarrow$ $Ar_2^*(^{1,3}\Sigma_u^+) + Ar$ | $k_1 \sim 1 \times 10^{-32}$ cm$^6$s$^{-1}$ | 300 K | [44–47] | $\sim$13 ns |
| (2) | $Ar_2^*(^{1,3}\Sigma_u^+) \rightarrow 2Ar + h\nu$ (VUV) | $\tau_2(^1\Sigma_u^+) = 4.2$ ns | 300 K | [48,49] | 4.2 ns |
| | | $\tau_2(^3\Sigma_u^+) = 3.0 - 3.2\,\mu$s | 300 K | [12,47–51] | 3.1 $\mu$s |
| (3) | $Ar^*(3p^54p^1) \rightarrow$ $Ar^*(3p^54s^1) + h\nu$(NIR) | $\tau_3 = 20$–40 ns | 300 K | [34,52,53] | |
| | | $\tau_3 < 100$ ns | 163 K | [54–56] | <100 ns |
| (4) | $Ar^*(3p^54s^1) + Xe \rightarrow Ar + Xe^*$ | $k_4 = (2–3) \times 10^{-10}$ cm$^3$s$^{-1}$ | 300 K | [13,57] | $\sim$1 ns |
| (5) | $Ar_2^*(^3\Sigma_u^+) + Xe \rightarrow$ $2Ar + Xe^*(^1P_1, {}^3P_0)$ | $k_5 \sim 5 \times 10^{-10}$ cm$^3$s$^{-1}$ | 300 K | [12,13,58] | $\sim$0.6 ms |
| (6) | $Ar^*(3p^54s^1) + N_2 \rightarrow Ar + N_2^*(C)$ | $k_6 \sim 1.5 \times 10^{-11}$ cm$^3$s$^{-1}$ | 300 K | [44,46] | |
| | | $k_6 = 3.6 \times 10^{-11}$ cm$^3$s$^{-1}$ | 300 K | [59] | 2.4 $\mu$s |
| | | $k_6 \geq 6.5 \times 10^{-9}$ cm$^3$s$^{-1}$ (?) | 87 K | [11] | $\leq$13 ns (?) |
| (7) | $Ar^*(3p^54s^1) + N_2 \rightarrow$ $Ar + N_2^*(C, B, A)$ | $k_7 \sim 3 \times 10^{-11}$ cm$^3$s$^{-1}$ | 300 K | [44,46] | |
| | | $k_7 = 3.6 \times 10^{-11}$ cm$^3$s$^{-1}$ | 300 K | [57,59] | |
| (8) | $N_2^*(C) \rightarrow$ $N_2^*(B) + h\nu$ (UV, 2nd pos. sys.) | $\tau_8 = 30$–40 ns | 300 K | [44,46,60] | 35 ns |
| (9) | $N_2^*(B) \rightarrow$ $N_2^*(A) + h\nu$ (NIR, 1st pos. sys.) | $\tau_9 \sim 9\,\mu$s | 300 K | [44] | $\sim$9 $\mu$s |
| | | | 119 K | [38] | |
| (10) | $N_2^*(C) + Ar \rightarrow N_2^*(B) + Ar$ | $k_{10} = 5.6 \times 10^{-13}$ cm$^3$s$^{-1}$ | 300 K | [44] | 21 ns |
| (11) | $N_2^*(B) + Ar \rightarrow N_2^*(A) + Ar$ | $k_{11} = 1.4 \times 10^{-14}$ cm$^3$s$^{-1}$ | 300 K | [44] | 0.8 $\mu$s |
| (12) | $N_2^*(C) + N_2 \rightarrow N_2 + N_2^*(B)$ | $k_{12} \sim 1 \times 10^{-11}$ cm$^3$s$^{-1}$ | 300 K | [44,60] | $\sim$8.6 $\mu$s |
| (13) | $N_2^*(B) + N_2 \rightarrow N_2 + N_2^*(A)$ | $k_{13} \sim 1 \times 10^{-11}$ cm$^3$s$^{-1}$ | 300 K | [44] | $\sim$8.6 $\mu$s |
| (14) | $Ar_2^*(^3\Sigma_u^+) + N_2 \rightarrow 2Ar + N_2^*(B)$ | $k_{14} \sim 3.3 \times 10^{-12}$ cm$^3$s$^{-1}$ | 300 K | [44,58] | $\sim$26 $\mu$s |
| | Liquid Ar + Xe (1000 ppm) + $N_2$ (50 ppm) | | | | |
| (15) | $Ar^*(n = 1, {}^2P_{1/2,3/2}) + Ar \rightarrow$ $Ar_2^*(^{1,3}\Sigma_u^+)$ | $\tau_{15} = 6$ ps | 87 K | [25,61] | 6 ps |
| (16) | $Ar_2^*(^{1,3}\Sigma_u^+) \rightarrow 2Ar + h\nu$ (VUV) | $\tau_{16}(^1\Sigma_u^+) = 7$ ns | 87 K | [8–10,62] | 7 ns |
| | | $\tau_{16}(^3\Sigma_u^+) = 1.6\,\mu$s | | | 1.6 $\mu$s |
| (17) | $Ar_2^*(^3\Sigma_u^+) + Xe \rightarrow$ $2Ar + Xe^*(n = 1, 2, {}^2P_{3/2})$ | $k_{17}(^3\Sigma_u^+) \sim$ $(0.8 - 1) \times 10^{-11}$ cm$^3$s$^{-1}$ | 87 K | [17–19] | $\sim$5.3 ns |
| | | $\tau_{17}(^3\Sigma_u^+) < 90$ ns | 87 K | [18,20] | <90 ns |
| | | $k_{17}(^1\Sigma_u^+) \sim 3.3 \times 10^{-11}$ cm$^3$s$^{-1}$ | 87 K | [19] | $\sim$1.4 ns |
| (18) | $Xe^*(n = 1, 2, {}^2P_{3/2}) + Ar \rightarrow ArXe^*$ | Immediate trapping | 87 K | [19] | |
| (19) | $ArXe^* + Xe \rightarrow Ar + Xe_2^*(^{1,3}\Sigma_u^+)$ | $\tau_{19} \leq 20$ ns | 87 K | [20] | $\leq$20 ns |
| (20) | $Xe^*(n = 1, 2, {}^2P_{3/2}) + Xe \rightarrow$ $Xe_2^*(^{1,3}\Sigma_u^+)$ | – | 87 K | [16] | |
| (21) | $Xe_2^*(^{1,3}\Sigma_u^+) \rightarrow 2Xe + h\nu$ (UV) | $\tau_{21}(^1\Sigma_u^+) = 4.3$ ns | 165 K | [9,62] | 4.3 ns |
| | | $\tau_{21}(^3\Sigma_u^+) = 22$ ns | 165 K | | 22 ns |
| (22) | $Xe^*(n = 2, {}^2P_{3/2}) \rightarrow$ $Xe^*(n = 1, {}^2P_{3/2}) + h\nu$ (NIR) | $\tau_{22} < 170$ ns | 87 K | [21,22] | <170 ns |
| | Reactions (17)–(21) in total ($\tau_{17} + \tau_{19}$): | | | | $\leq$110 ns |
| (23) | $Ar_2^*(^3\Sigma_u^+) + N_2 \rightarrow 2Ar + N_2^*(B)$ | $k_{23} = 3.8 \times 10^{-12}$ cm$^3$s$^{-1}$ | 87 K | [24,25] | 250 ns |
| (24) | $ArXe^* + N_2 \rightarrow Ar + Xe + N_2^*(B, A)$ | – | 87 K | | |
| (25) | $Xe_2^*(^3\Sigma_u^+) + N_2 \rightarrow 2Xe + N_2^*(B, A)$ | – | 87 K | | |

## 3.2. Electroluminescence due to excimer emission

In ordinary (excimer) EL, the electrons accelerated by the electric field excite the atoms, which in turn produce the excimers (excited molecules) in three-body collisions. The excimers are produced in either a singlet $Ar_2^*(\,^1\Sigma_u^+)$ or a triplet $Ar_2^*(\,^3\Sigma_u^+)$ state. These states decay with photon emission in the VUV, peaked at 128 nm for Ar and 175 nm for Xe (see Figures 3 and 4 and reactions (1) and (2) in Table 1):

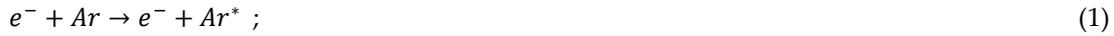

$$e^- + Ar \rightarrow e^- + Ar^* \;\; ; \tag{1}$$

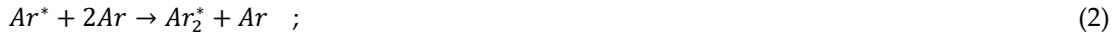

$$Ar^* + 2Ar \rightarrow Ar_2^* + Ar \;\; ; \tag{2}$$

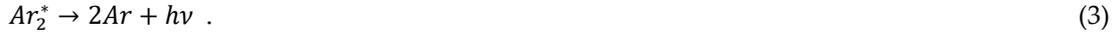

$$Ar_2^* \rightarrow 2Ar + h\nu \; . \tag{3}$$

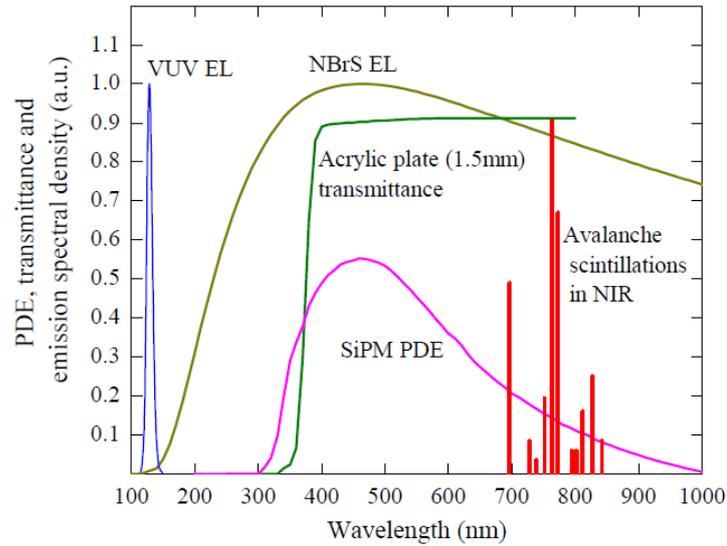

**Figure 4.** Photon emission spectra in gaseous Ar due to ordinary scintillations in the VUV measured in [8], NBrS EL at 8.3 Td theoretically calculated in [9] and avalanche scintillations in the NIR measured in [10,11]. Also shown are the Photon Detection Efficiency (PDE) of SiPM (MPPC 13360-6050PE (Hamamatsu)) at overvoltage of 5.6 V and the transmittance of the acrylic plate (1.5 mm thick) [12].

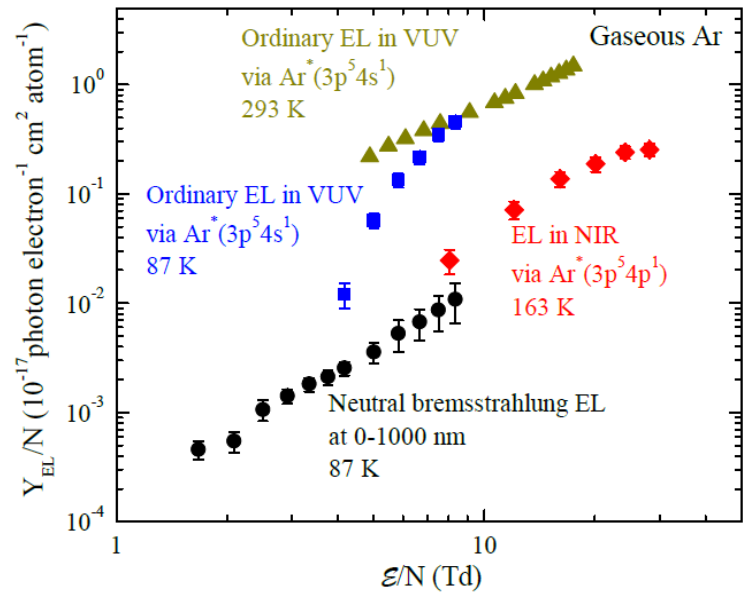

**Figure 5.** Summary of experimental data on reduced EL yield in gaseous Ar for all known electroluminescence (EL) mechanisms: for NBrS EL below 1000 nm, measured in [9,13] at 87 K; for ordinary EL in the VUV, due to excimer emission going via $Ar^*(3p^54s^1)$ excited states, measured in [9,13] at 87 K and in [14] at 293 K; for EL in the NIR due to atomic transitions going via $Ar^*(3p^54p^1)$ excited states measured in [15] at 163 K.

The singlet and triplet excimer states are responsible respectively for the fast and slow emission components observed in proportional EL. Their time constants are of 4.2 ns and 3.1 μs respectively: see Table 1 and references therein. Ordinary EL has a threshold in the electric field, of about 4 Td (see Figure 5), defined by the energy threshold of 11.55 eV of Ar atom excitation in Equation (1).

Ordinary proportional EL has two characteristic properties: a threshold and linear field dependence (far enough from the threshold). For Xe, this can be seen from Figure 6 presenting compilation of the experimental data on EL yield by 2007: the results of different groups are close, with the exception of early works.

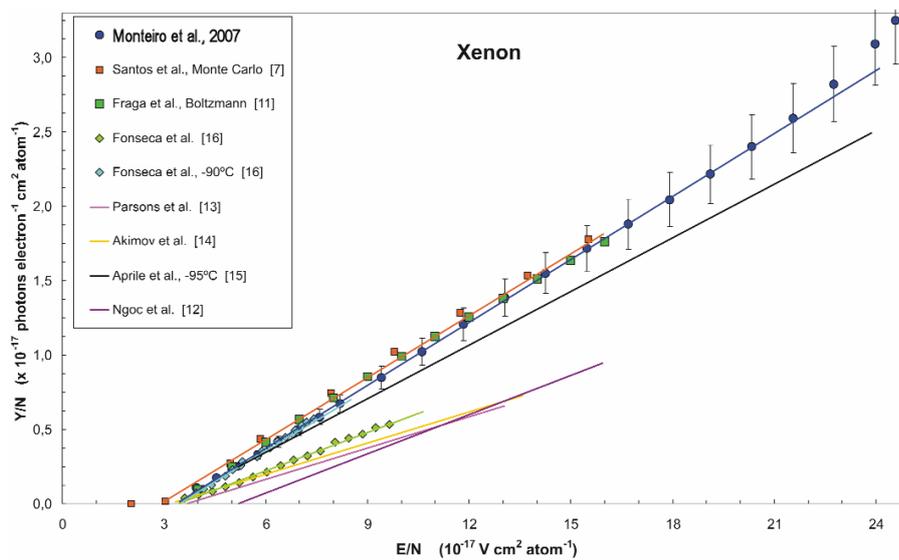

**Figure 6.** Reduced EL yields for ordinary (excimer) EL in Xe measured by different groups, as a function of the reduced electric field [17]. The references in the figure are those of [17].

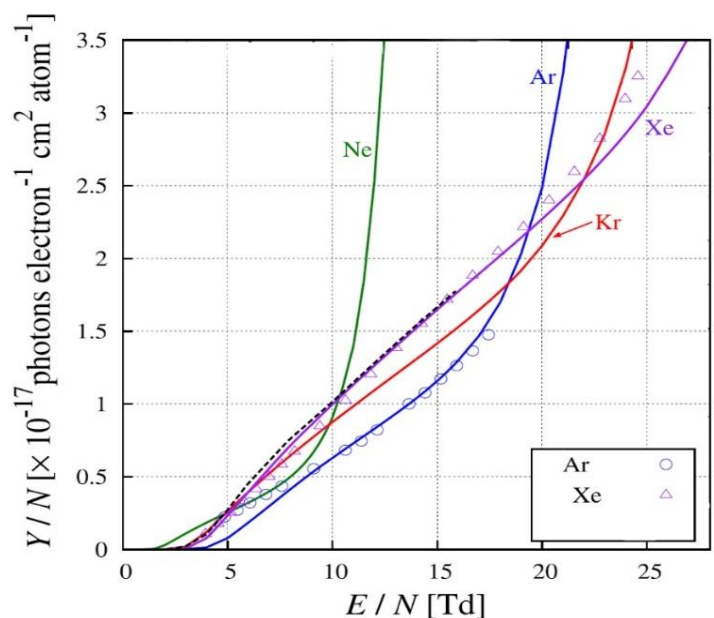

**Figure 7.** Measured reduced EL yield for ordinary (excimer) EL at room temperature for Ar and Xe (data points), as a function of the reduced electric field, compared to Monte Carlo simulation data (curves), for Ne, Ar, Kr and Xe [16].

The absolute photon yields of ordinary (excimer) EL were theoretically calculated for Ne, Ar, Kr and Xe in [16] using Monte Carlo simulation in microscopic approach: see Figure 7. One can see from the figure that well above the EL threshold, the theoretical predictions are in rather good agreement with the yields measured in experiment at room temperature, for Ar and Xe. Extrapolating the theoretical EL yield, the EL threshold was determined as 1.5, 4.1, 2.6 and 2.9 Td for Ne, Ar, Kr and Xe respectively [16].

Far enough from the threshold, the EL yield can be parametrized by a linear function, indicating the proportionality of the EL yield to the electric field:

$$Y/N \ [10^{-17} \text{photon electron}^{-1} \text{ cm}^2 \text{ atom}^{-1}] = a \cdot \ E/N \ - b \quad, \tag{4}$$

where reduced electric field $E/N$ is given in Td ($10^{-17}$ V cm² atom⁻¹). This equation is universally valid for all temperatures and gas densities. Parameter "$b$" defines the electric field threshold for proportional EL, while parameter "$a$" describes the light amplification factor, namely the number of photons emitted by drifting electron per applied voltage across the EL gap. For Ar, Kr and Xe this parameter was measured to be 81 [14], 120 [18] and 140 [17] photon/kV respectively.

The parametrization of the experimental data by linear field dependence was presented in [14] and [17] for Ar and Xe respectively:

$$Y/N = 0.081 \cdot \ E/N \ - 0.190 \ , \text{for Ar};$$
$$Y/N = 0.140 \cdot \ E/N \ - 0.474 \ , \text{for Xe}. \tag{5}$$

However, near the EL threshold this parametrization gives overestimated EL yield values, due to contribution of neutral bremsstrahlung electroluminescence (see section 3.3). In this case the experimental data relevant to ordinary EL should be used directly. In particular, in two-phase Ar at 87 K and 1.0 atm, the number of photons in the VUV emitted by drifting electron per 1 cm, deduced from experimental data of Figure 5, is 345 and 43 at electric field of 8 and 4.6 Td respectively, the latter corresponding to the nominal operation field of the DarkSide experiment.

### 3.3. Electroluminescence due neutral bremsstrahlung effect

In addition to the ordinary EL mechanism, there is a concurrent EL mechanism, based on bremsstrahlung of drifting electrons scattered on neutral atoms, so-called neutral bremsstrahlung (NBrS) [9,13,19]:

$$e^- + Ar \rightarrow e^- + Ar + h\nu \ . \tag{6}$$

The development of the mechanism is described in detail in [9].

The NBrS differential cross section is proportional to elastic cross section ($\sigma_{el}(E)$) of electron-atom scattering [9]:

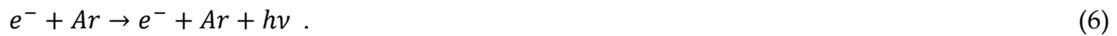

$$\frac{d\sigma}{d\nu} = \frac{8}{3} \frac{r_e}{c} \frac{1}{h\nu} \left(\frac{E-h\nu}{E}\right)^{1/2} [(E-h\nu) \cdot \sigma_{el}(E) + E \cdot \sigma_{el}(E-h\nu)] \quad . \tag{7}$$

Here $r_e = e^2/m_e c^2$ is the classical electron radius, $c = \nu\lambda$ is the speed of light, $E$ is the initial electron energy. For electron-atom scattering cross sections in Ar, see Figure 8.

Consequently, the NBrS EL yield is predicted to be maximal for electron energies of the order of 10 eV where the elastic cross section has a maximum, thus corresponding to the electric fields of 4-5 Td. This is confirmed in Figure 9 showing the absolute yield of NBrS EL theoretically calculated in [9], in comparison with that of ordinary EL.

NBrS EL has a continuous emission spectrum, extending from the UV (200 nm) to the visible and NIR range (1000 nm) and even to the radio frequency range [20]: see Figure 4. From Figures 4 and 9 one can see that NBrS EL, albeit being significantly weaker than ordinary EL above the Ar excitation threshold, has no threshold in electric field, in contrast to ordinary EL, and thus dominates below the threshold. Accordingly, the NBrS effect can explain two remarkable properties of proportional EL observed in experiment [9,13,19]: that of the photon emission below the Ar excitation threshold and that of the substantial contribution of the non-VUV spectral component above the threshold.

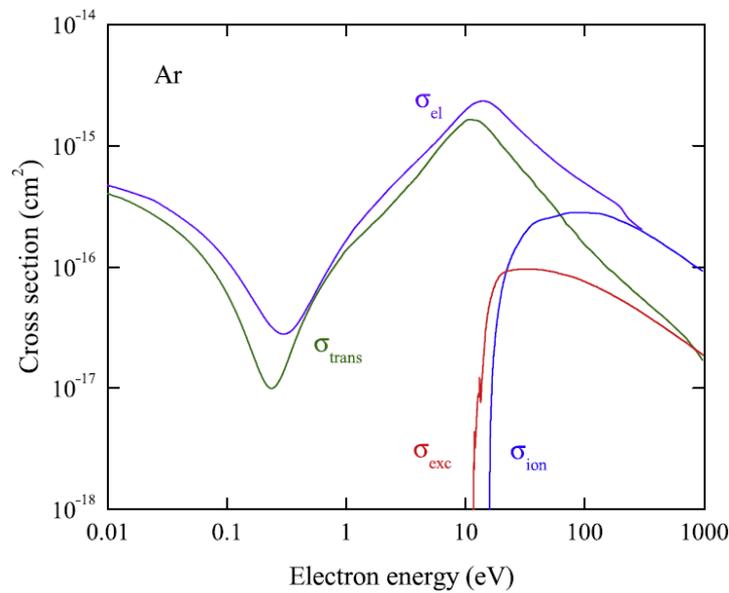

**Figure 8.** Electron scattering cross sections in Ar [9]: elastic ($\sigma_{el}$), momentum-transfer ($\sigma_{trans}$), excitation ($\sigma_{exc}$) and ionization ($\sigma_{ion}$).

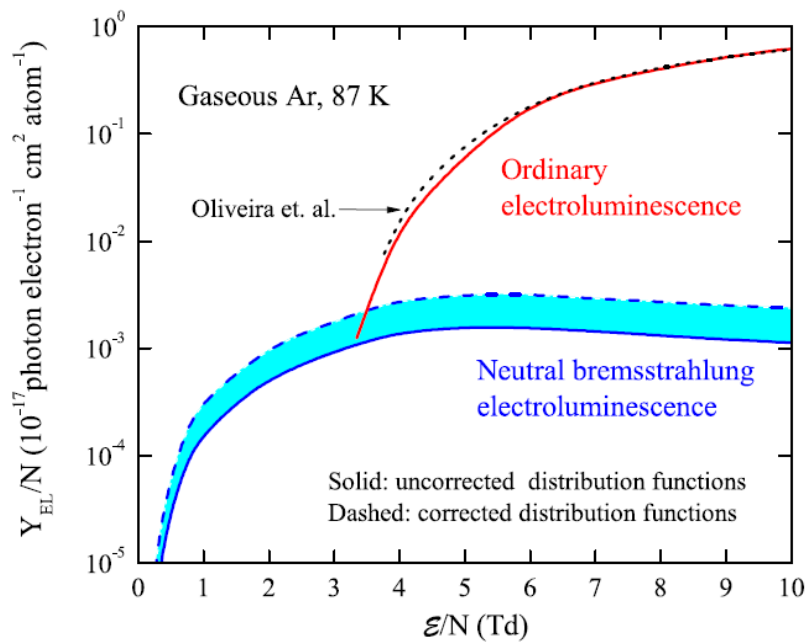

**Figure 9.** Reduced EL yield in gaseous Ar as a function of the reduced electric field for neutral bremsstrahlung EL (below 1000 nm) [9], calculated theoretically in [9] and compared to ordinary (excimer) EL calculated in [16].

At lower electric fields, below 4 Td corresponding to Ar excitation threshold, the NBrS theory developed in [9] correctly predicts the absolute value of the EL yield. This is seen when comparing the experimental data on EL yields in Figure 5 to those of the theory in Figure 9 and when comparing the experimental and theoretical photon emission spectra in Figure 10. On the other hand, at higher fields (above 5 Td), the experimental EL yields quickly diverge from those of the theory, exceeding those in the UV spectral range (below 400 nm). In [9] this discrepancy was hypothesized to be explained by the contribution of electron scattering on sub-excitation Feshbach resonance (going via intermediate negative ion state $Ar^-(3p^54s^2)$), which might be accompanied by the enhanced photon emission [21].

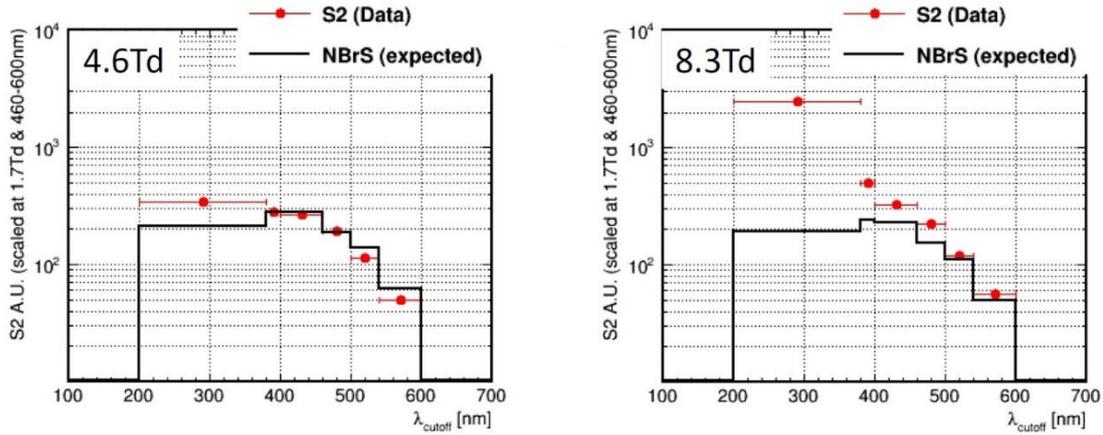

**Figure 10**. Photon emission spectra of proportional EL in Ar in the UV and visible range [19], measured at room temperature in [19] (data points) in comparison with those of the NBrS EL theory [9] (lines), at reduced electric field of 4.6 and 8.3 Td.

Near the threshold of ordinary EL, namely in the field range of 4-6 Td, it becomes important to take into account the possible contribution of NBrS EL, to correctly measure the EL yield of ordinary EL. Note that just in this field region, namely nominally at 4.6 Td, the DarkSide dark matter search experiment operates [22]. In [9,13] the NBrS contribution was measured, using dedicated spectral devices insensitive to the VUV, and appropriately subtracted from the data obtained with VUV-sensitive devices. The resulted EL yield of ordinary EL, shown in Figure 11, is in good agreement with the theory. On the other hand, the experimental data of [14] in Figure 11 have an excess over those of [9] and theory [16], below 6 Td. It would be logical to explain this discrepancy by the NBrS contribution that was not taken into account in [14].

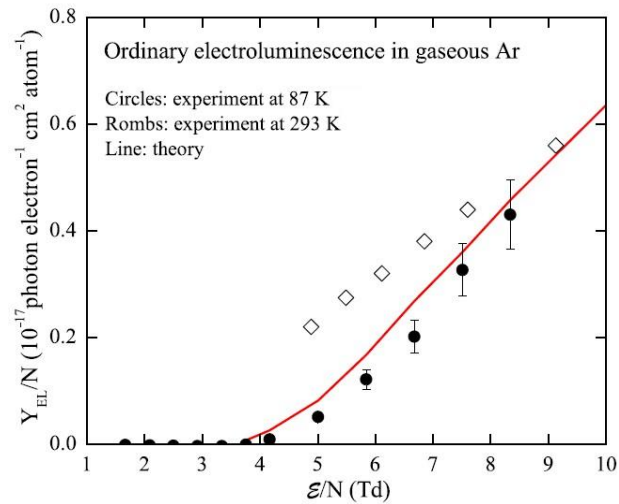

**Figure 11.** Reduced EL yield in gaseous Ar as a function of the reduced electric field, for ordinary EL (due to excimer emission in the VUV) near its threshold [9], measured at 87 K [9] and at 293 K [14]. The theoretical calculation of [16] is also shown.

### 3.4. Electroluminescence due to atomic transitions in the NIR

At higher electric fields, above 8 Td, another "non-standard" EL mechanism comes into force, namely that of EL in the NIR due to transitions between excited atomic states [4,7,10,11,15,23]. Here the higher atomic levels $Ar^*(3p^5 4p^1)$ are excited; the transitions between those and the lower excited levels are responsible for the fast emission in the NIR, at $700 - 850$ nm (see Figures 3 and reactions (3) in Table 1):

$$e^- + Ar \rightarrow e^- + Ar^*(3p^54p^1) \; ; \tag{8}$$
$$Ar^*(3p^54p^1) \rightarrow Ar^*(3p^54s^1) + h\nu \; . \tag{9}$$

Atomic EL in the NIR has a line emission spectrum: see Figure 4. In addition, it has higher threshold in electric field than ordinary EL (see Figure 12). The absolute yield of atomic EL in the NIR was measured in [15,24]; it was well described by the theory using microscopic approach [23]: see Figure 12. It should be remarked that atomic emission in the NIR is particularly noticeable at even higher fields, above 30 Td, where the avalanche multiplication of the electrons takes place, accompanied by corresponding secondary scintillations: by so-called "avalanche scintillations" [11,25].

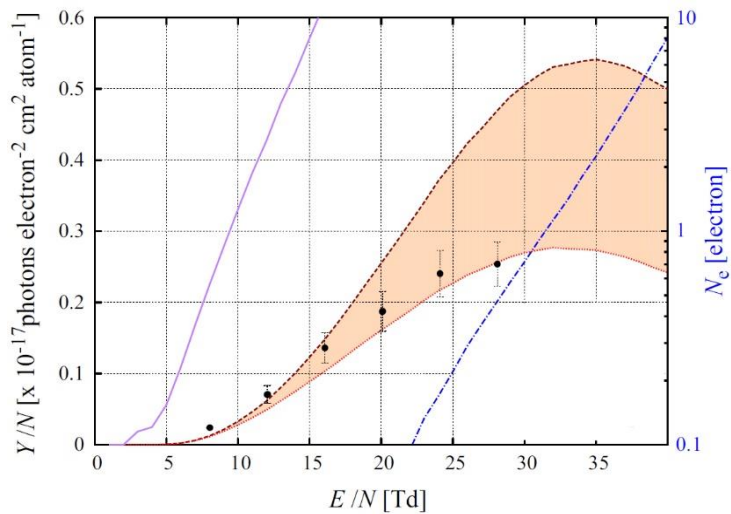

**Figure 12.** Reduced EL yield in gaseous Ar as a function of the reduced electric field, for atomic EL in the NIR due to atomic transitions going via $Ar^*(3p^54p^1)$ excited states [23], measured in [15] at 163 K (data points) and theoretically calculated in [23] (hatched area between two curves). For comparison, that for ordinary EL in the VUV going via $Ar^*(3p^54s^1)$ (solid curve) and the number of secondary electrons due to electron avalanching in 2 mm gap (dashed-dot curve, right scale), theoretically calculated in [23], are shown.

### 3.5. Concepts of light signal amplification

Ordinary EL in Ar and Xe results in two "standard" concepts of light signal amplification in two-phase detectors, depicted in Figure 1. In the first concept used in Ar, the EL gap is optically readout by the SiPM or PMT matrix indirectly, via a wavelength shifter (WLS), typically TPB. This is because the typical cryogenic PMTs and SiPMs are not sensitive to ordinary EL of Ar, at 128 nm.

In the second concept used in Xe, the EL gap is optically read out directly by the PMT matrix, provided that the typical cryogenic PMTs has quartz windows and thus become sensitive to ordinary EL of Xe, at 175 nm.

The concept of direct optical readout (Figure 1 (right)), used for Xe, can also apply to Ar to detect NBrS EL in the non-VUV range. Such an alternative readout concept has the advantage of doing without WLS, in two-phase Ar detectors. This may lead to more stable operation due to avoiding the problems of WLS degradation and its dissolving in liquid [26], as well as that of WLS peeling off from the substrate. Such a concept was realized in [12]. It might be considered as backup solutions for large-scale two-phase Ar detectors, in case the problem of WLS instability cannot be resolved.

## 4. Charge signal amplification in the gas phase of two-phase detector, using electron avalanching

### 4.1. Charge signal amplification concepts at cryogenic temperatures

Over the past two decades, there has been a growing interest in so-called "Cryogenic Avalanche Detectors": see review [4]. In the wide sense these define a class of noble-gas detectors operated at cryogenic temperatures with electron avalanching (electron multiplication) performed directly in the detection medium, the latter being in a gaseous, liquid or two-phase state, provided by electron impact ionization reaction:

$$e^- + Ar \rightarrow 2e^- + Ar^+ \ . \tag{10}$$

Earlier attempts to obtain high and stable electron avalanching directly in noble gases and liquids at cryogenic temperatures, using "open-geometry" multipliers, have not been very successful: rather low gains (≤10) were observed in liquid Xe [27,28,29,30] and Ar [31] and low gains (≤100) in gaseous Ar [32] and He [33] at cryogenic temperatures, using wire, needle or micro-strip proportional counters. Moreover, two-phase detectors with wire chamber readout, which initially seemed to solve the problem, turned out to have unstable operation in the avalanche mode due to vapour condensation on wire electrodes [34].

The problem of electron avalanching in cryogenic noble-gas detectors was solved in 2003 [35] after introduction of cryogenic gaseous and two-phase detectors with Gas Electron Multiplier (GEM) readout. GEM [36] and thick GEM (THGEM, [37]) belongs to the class of Micro-Pattern Gas Detectors (MPGDs). GEM is a thin insulating film, metal clad on both sides, perforated by a matrix of micro-holes, in which gas amplification occurs under the voltage applied across the film. The typical GEM geometrical parameters are the following: dielectric (Kapton) thickness is 50 μm, hole pitch - 140 μm, hole diameter on metal - 70 μm. THGEM is a similar, though more robust structure with about ten-fold expanded dimensions. In particular, the typical THGEM geometrical parameters are the following: t/p/d/h=0.4/0.9/0.5/0.1 mm. Here t/p/d/h denotes "dielectric thickness/hole pitch/hole diameter/hole rim width".

Contrary to wire chambers and other "open geometry" gaseous multipliers, cascaded GEMs and THGEMs have a unique ability to operate in dense noble gases at high gains [38], including at cryogenic temperatures and in the two-phase mode [39]. Consequently at present, the basic idea of Cryogenic Avalanche Detectors (CRADs) in the narrow sense is that of the combination of MPGDs with cryogenic noble-gas detectors, operated in a gaseous, liquid or two-phase mode.

The original CRAD concept [35] is the following (see Figure 2): electron avalanching at cryogenic temperatures is performed in pure noble gases using GEM multiplier, either in a gaseous or two-phase mode. In the latter case the conventional two-phase electron-emission detector is provided with charge signal amplification using electron avalanching in the gas phase: the primary ionization electrons produced in the liquid are emitted into the gas phase by an electric field, where they are multiplied in saturated vapour using cascaded GEM multiplier. The proof of principle of this concept was demonstrated in 2003-2006 in two-phase Ar, Kr and Xe [35,40] at appropriate cryogenic temperatures (at around 87 K, 120 K and 165 K, respectively) and in gaseous He and Ne at lower temperatures, down to 2.6 K [41,42].

Later on, the original CRAD concept was elaborated: it was suggested to provide CRADs with new features. There are more than a dozen of different concepts with charge and light amplification in two-phase and liquid detectors developed by different groups since 2003: their gallery by 2012 is shown in Figure 13. For their detailed description, we refer to review [4] and appropriate references listed in the figure: [15,25,35,40,43,44,45,46,47,48,49,50,51,52,53,54]. The references to such concepts and their realization after 2012 are as follows: [24,55,56,57,58,59,60,61,62].

In sections 4-6, we confine ourselves to the discussion of the most developed and successfully realized concepts, as well as of the physical effects and specific technology underlying them.

**Figure 13.** Gallery of concepts of charge signal amplification in two-phase detectors, using electron avalanching in the gas phase and charge or optical readout, by 2012 in order of introduction.

*4.2. GEM operation in pure noble gases at cryogenic temperatures*

In most CRAD concepts, the MPGD multiplier should be able to operate at high gains in dense pure noble gases, in particular in saturated vapour and at cryogenic temperatures. Fortunately, unlike open-geometry gaseous multipliers (e.g. wire chambers), cascaded GEMs and THGEMs permit attaining high (≥10³) charge gains in "pure" noble gases at atmospheric pressure, presumably due to considerably reduced photon-feedback effects. At room temperature, this remarkable property was discovered first for Ar [38,63] and then for other noble gases [4,39,64,65], using GEM multipliers. For operation of other MPGD multipliers in "pure" noble gases at room temperature, we refer to review [4].

It should be remarked that the stable operation of GEMs at cryogenic temperatures is a non-trivial fact, since the electrical resistance of their dielectric materials considerably increases with the temperature decrease, i.e. by an order of magnitude per 35 degrees for Kapton GEMs [66], which might result in strong charging-up effects within the holes. Fortunately, the latter did not happen: the charging-up effects for GEMs have not been observed. The GEM performance in gases at cryogenic temperatures was found to be generally independent of temperature down to ~100 K, at a given gas density [66]: stable and high-gain GEM operation was observed in all noble gases and in their mixtures with selected molecular additives that do not freeze in a wide temperature range ($CH_4$, $N_2$ and $H_2$). This is seen from Figure 14 (left):  rather high triple-GEM gains were reached at cryogenic temperatures, exceeding $10^4$ in Ar, Kr and Xe+$CH_4$.

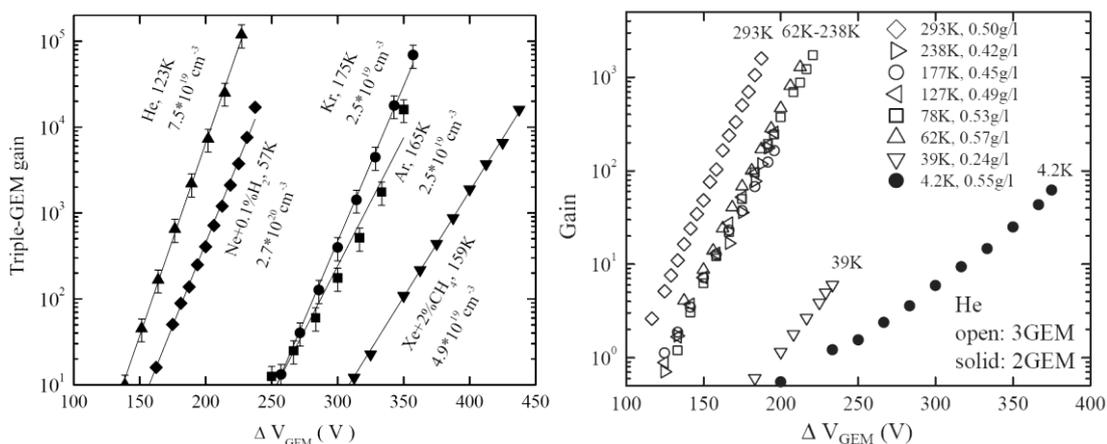

**Figure 14.** Gain characteristics of GEM multipliers at cryogenic temperatures [4]. Left: in gaseous He, Ar and Kr, in the Penning mixture Ne+0.1%$H_2$ (its density corresponding to saturated Ne vapour in the two-phase mode) and in Xe+2%$CH_4$. Right: in gaseous He at low temperatures (down to 4.2 K).  The appropriate temperatures and densities are indicated.  In He at 39 K and 4.2 K, the maximum gains were limited by discharges, while at other temperatures and in other gases the discharge limit was not reached. The multiplier active area was 2.8x2.8 cm².

It is amazing that GEMs were able to operate in electron avalanching mode at even lower temperatures, down to 2.6 K in gaseous He [41,42]. On the other hand, high GEM gains observed in He and Ne above 77 K were reported to be due to the Penning effect in uncontrolled (≥$10^{-5}$) impurities (i.e. $N_2$) which froze out at lower temperatures, resulting in the considerable gain drop at temperatures below 40 K: see Figure 14 (right). A solution to the gain drop problem at lower temperatures was found in [41]: Ne and He can form high-gain Penning mixtures with $H_2$ at temperatures down to ~10 K. This is seen from Figure 14 (left): triple-GEM gains exceeding $10^4$ were obtained at 57 K in the Penning mixture Ne+0.1%$H_2$, its density corresponding to that of saturated Ne vapour in the two-phase mode.  Unfortunately, this does not work for two-phase He, due to the very low $H_2$ vapour pressure at 4.2 K.

*4.3. Two-phase detectors with GEM multipliers*

Regarding GEM operation in two-phase detectors, most promising results were obtained for two-phase Ar detectors (see Figure 15): the charge gain of 5000 with triple-GEM readout was routinely attained [40] and the stable operation for tens of hours at this gain was demonstrated [67]. This maximum gain value should be compared to that of 600 and 200 obtained in GEM-based two-phase Kr and Xe detectors respectively: see Figure 15.

In this review the charge gain is defined as the ratio of the output anode charge of the GEM or THGEM multiplier incorporated into the two-phase detector to the input "primary" charge, i.e. to that of prior to multiplication measured in special calibration runs.

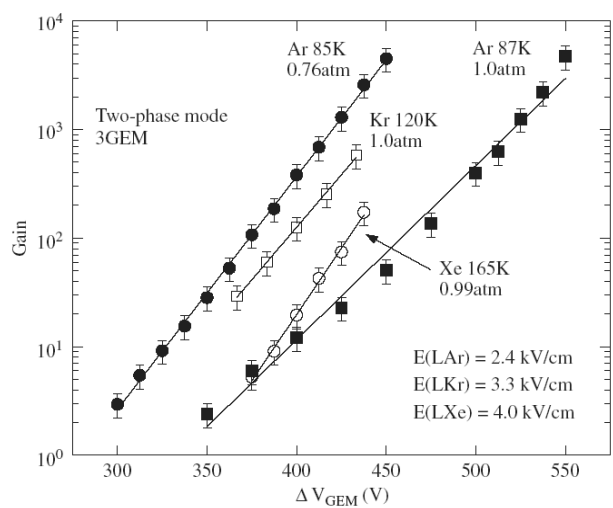

**Figure 15.** Gain characteristics in two-phase Ar, Kr and Xe detectors with triple-GEM multiplier charge readout [40]. The appropriate temperatures, pressures and electric fields in the liquids are indicated. The multiplier active area was 2.8x2.8 cm²; the maximum gains were limited by discharges.

The successful performance of two-phase Ar detectors with triple-GEM charge readout is illustrated in Figures 16 and 17. In particular, the high triple-GEM gain provided the operation of the two-phase detector in single-electron counting mode (Figure 16) [68]. It should be remarked that though the single-electron spectra are well separated from electronic noise at gains exceeding 5000, they are described by an exponential function (rather than by a peaked function). The latter is generally a rule for gaseous multipliers operated in proportional mode, due to intrinsically considerable fluctuations of the avalanche size. Consequently, the single- and double-electron events can hardly be distinguished in two-phase Ar detectors with GEM (or THGEM) charge readout. For that, the combination with PMT- or SiPM-based optical readout should be used, as suggested in [15,24]. Accordingly, the function of GEM- or THGEM-based charge readout would be to provide superior spatial resolution.

Figure 17 demonstrates the unique ability of the two-phase Ar detector with GEM charge readout to observe directly and simultaneously the fast and slow components of electron emission through the liquid-gas interface [69]. The fast component is due to the hot electrons, heated by an electric field, while the slow component is due to the electrons reflected from the potential barrier and then thermalized [4]. Note that the two-phase Ar detector with optical readout of the EL gap cannot provide such an ability, due interference with the slow component of excimer (VUV) emission. It should be also emphasized that the slow component has never been observed in two-phase Kr and Xe systems, presumably due to higher potential barrier at the interface as compared to Ar. And vice versa, the fast component has never been observed in two-phase He and Ne systems since the electrons in liquid He and Ne are localized in the bubbles and thus cannot be heated by the electric field. Consequently, the two-phase Ar system is unique in this sense.

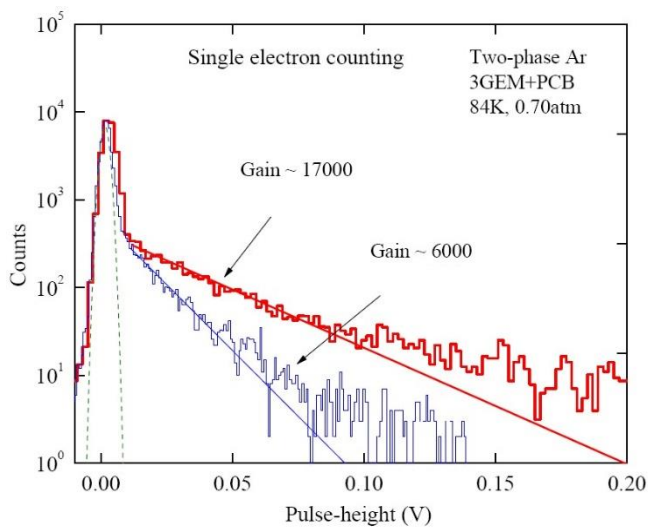

**Figure 16.** Pulse-height spectra in two-phase Ar detector with triple-GEM multiplier charge readout, operated in single electron counting mode with external trigger, at charge gains of 6000 and 17000 [68]. The electronic noise spectrum (dashed line), corresponding to the Equivalent Noise Charge of $\sigma$=940 e⁻, is also shown. The multiplier active area was 2.8x2.8 cm². The results were obtained for specially selected GEM foils resistant to discharges, to reach the highest gain. Note that the stable maximum gain of typical triple-GEM multipliers in two-phase Ar was somewhat lower: of about 5000.

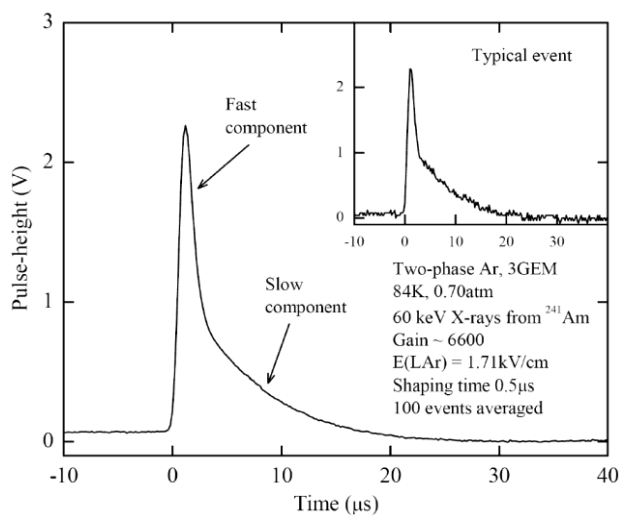

**Figure 17.** Typical anode signals in two-phase Ar detector with triple-GEM multiplier charge readout at emission electric field in the liquid of 1.7 kV/cm [69]. The fast and slow components of electron emission through liquid-gas interface are distinctly seen.

It should be remarked that all the gain data of this section were obtained with GEMs with relatively small active area, of 2.8x2.8 cm². Unfortunately, as we will see later, the maximum gain decreases with the multiplier active area.

### 4.4. Two-phase detectors with THGEM multipliers

In two-phase detectors with THGEM charge readout, the maximum gains are comparable to those with GEM readout: see Figures 18 and 19(left). Gains as high as 3000 [48] and 600 [70] were obtained in two-phase detectors in Ar and Xe respectively, with double-THGEM multipliers having active area of 2.5×2.5 cm².

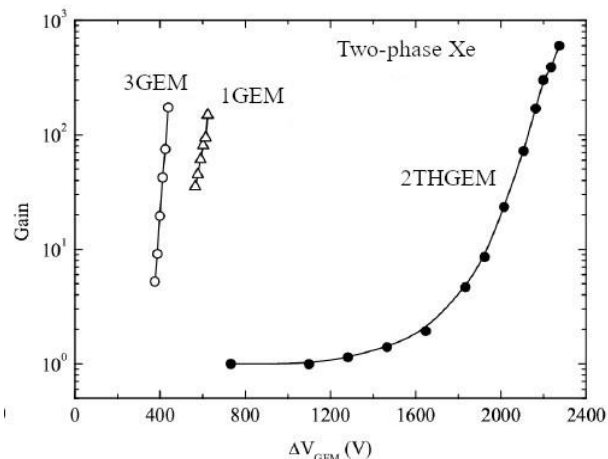

**Figure 18.** Gain characteristics of double-THGEM multipliers in two-phase Xe detector for THGEM active area of 2.5x2.5 cm² [70]. Gain characteristics of triple-GEM [40] and single-GEM [71] multipliers (of similar active area) are shown for comparison. Here the maximum gains were limited by discharges.

However for larger active area, of 10×10 cm², the maximum gain in two-phase Ar detector decreased three-fold: down to about 1000 for double-THGEM multiplier [70] and 100-200 for single-THGEM multiplier [72,73]: see Figures 19 (left) and 20 (right). This could result from the larger number of holes, implying a larger discharge probability on defects. Note that the nominal electric field in Figure 20, defined as the THGEM voltage divided by its dielectric thickness, is higher than the real electric field in the THGEM hole.

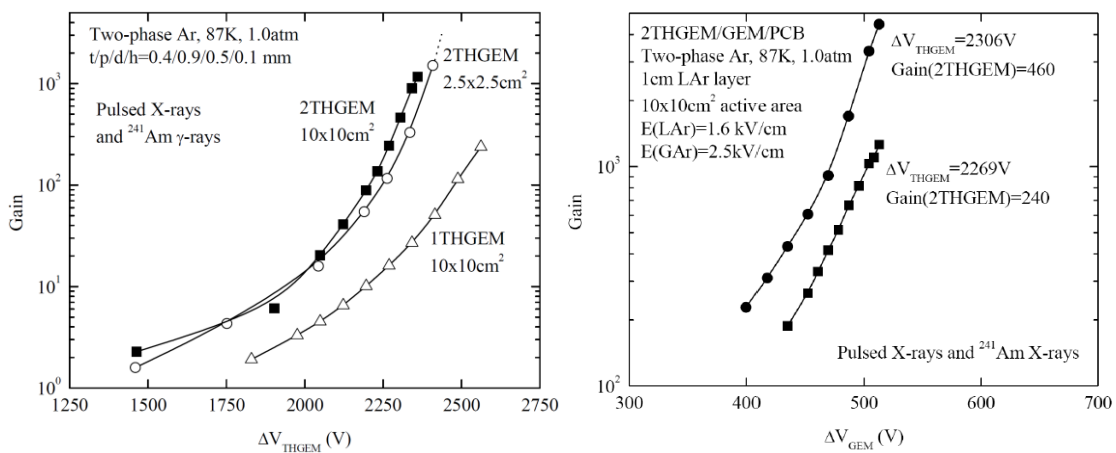

**Figure 19.** Left: gain characteristic of double-THGEM multipliers in two-phase Ar detector for THGEM active area of 10x10 cm² (maximum gain was limited by discharges) compared to that of 2.5x2.5 cm² (maximum gain was not reached) [72]. Also shown that of single-THGEM multiplier (maximum gain was limited by discharges). Right: gain characteristics of hybrid 3-stage 2THGEM/GEM multiplier in two-phase Ar detector for active area of 10x10 cm² [72]. Shown is the overall multiplier gain as a function of the voltage across GEM, at two fixed voltages across each THGEM, i.e. at two values of 2THGEM gain indicated in the figure. Here the maximum gains were limited by discharges.

The maximum gains of 100-1000 are obviously not sufficient for charge readout in a single-electron counting mode. Nevertheless, the gain of about 80, obtained in two-phase Ar detector with single-THGEM having 2D readout [73], permitted to obtain track images, demonstrating the excellent imaging capability of two-phase Ar detectors with THGEM multiplier charge readout even at such a moderate gain. This technique is of particular importance for giant liquid Ar detectors for long-baseline neutrino experiments. At present, the R&D of this technique has continued as part of the ArDM and DUNE experiments [74,75,76]. It should be remarked that in that community another terminology is used, namely "dual-phase TPC" and "Large Electron Multiplier" (LEM).

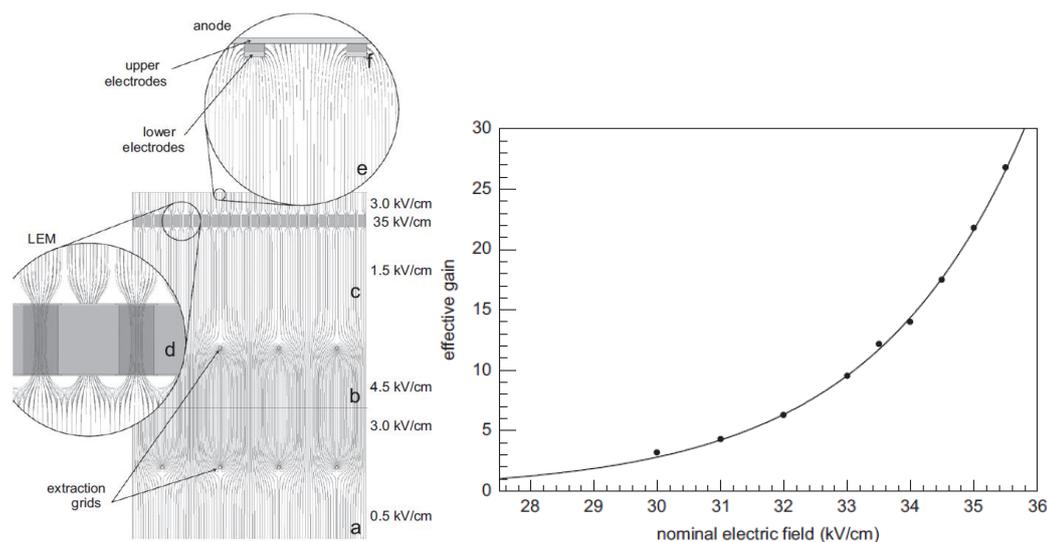

**Figure 20.** Left: electric field configuration in two-phase detector with THGEM (LEM) 2D readout [73]. Right: "Effective" gain characteristic for single-THGEM multiplier charge 2D readout in two-phase Ar detector, of an active area of 10x10 cm² [73], as a function of the nominal electric field in THGEM. The maximum gain was limited by discharges. The "effective" gain value should be multiplied by about a factor of 3, to be normalized to the gain definition of the present review (see details in [4]).

The further reduction of the maximum charge gain in two-phase Ar detector when increasing the active area up to 40x40 and 50x50 cm², was reported in [74] and [76] respectively. For the latter, the THGEM multiplier could not operate at nominal electric fields higher than 28 kV/cm, i.e. at charge gain higher than 3 according to Figure 20 (right). This is the real problem; it is discussed below.

An interesting way to increase the maximum charge gain in two-phase Ar CRADs was suggested in [72], namely that using a hybrid 3-stage 2THGEM/GEM multiplier: see Figure 19 (right). The idea was that the avalanche charge from a THGEM hole would be shared between several GEM holes. This might reduce the overall avalanche charge density, believed to be responsible for the discharge development, and thus might increase the discharge voltage. The maximum gain as high as 5000 was attained for 10x10 cm² active area, in practical configuration when the last multiplier stage was followed by a Printed Circuit Board (PCB) anode, convenient for readout electronics coupling. This is almost enough for stable operation in single-electron counting mode with 2D readout, for which charge gains of 10⁴ are needed. It should be noted that the idea of hybrid multiplier, combining hole multipliers with different hole densities, can be realized without "fragile" GEMs: by cascading more robust THGEMs of different geometry, with sequential increasing hole densities and decreasing thicknesses.

### 4.5. Gain limit, gain stability and discharge-resistance problems in two-phase detectors with GEM and THGEM multipliers

Several problems of the performance of two-phase Ar detectors with GEM and THGEM charge readout remain unsolved. The first problem is that of the gain limit due to discharges. The obvious way to increase the gain would be adding an additional multiplication stage, i.e. up to overall 4 stages in the case of GEMs and 3 stages in the case of THGEMs. Also, it looks attractive to combine GEMs and THGEMs in a hybrid multiplier like that discussed above. Another way is an optical readout from THGEMs. Indeed, the THGEM charge gain of the order of 100 is still significant if combined with optical readout using SiPM matrices [55] or CCD cameras [60], and might be sufficient for track imaging and even for single-electron counting mode. These will be discussed in the following section.

The second problem is that of the resistance to electrical discharges of "standard" (thin Kapton) GEMs. It was observed [77] that when operating two-phase Ar detectors with triple-GEM multiplier at maximum gains (approaching 10⁴), the triple-GEM was not able to withstand electrical discharges on a long term: after several series of measurements, the maximum reachable gain decreased by several times. Obviously, this is due the low resistance of thin GEMs to discharges because of metal and/or carbon evaporation and their deposition on the insulator in the GEM holes. The solution of the problem might be switching to thicker THGEMs that were found to behave in more reliable way under discharges, or else combining THGEM and GEM multipliers or THGEM multipliers with increasing hole density, as discussed above.

There is another unsolved problem in two-phase CRAD performances. Namely, the performance of MPGD multipliers in two-phase Ar and Xe is not fully understood: not all multiplier types were able to operate with electron multiplication in saturated vapour. In two-phase Ar, while G10-based THGEM multipliers successfully operated for tens of hours with gains reaching a few thousands, others, namely RETHGEM and Kapton THGEMs, did not show stable multiplication in the two-phase mode [4,72]: they either did not multiply at all (with gain below 1) or showed unstable operation due to large gain variations. In addition, in [78] it was reported on the unsuccessful performance of the Micromegas multiplier in two-phase Xe: in half an hour the multiplication collapsed. The most rational explanation of these instabilities is the effect of vapour condensation within the THGEM holes or Micromegas mesh that prevents electron multiplication. The criteria for such a condensation are not yet clear. It could be that the specific properties of the holes and electrodes might play a role, i.e. the wetting capability depending on the electric-field non-uniformity and consequently on the MPGD geometry, as well as on the temperature gradients, the latter in turn depending on the electrode's heat conductivity.

*4.6. THGEM as interface grid in two-phase detectors*

In addition to the performance as charge multiplier, THGEM can be used as an interface grid immersed in the liquid (see Figure 1), acting as an effective electron transmission electrode defining the electron emission and EL regions of the two-phase detector [79]. Examples of such THGEM electrodes used in practice [9,12] are shown in Figure 21, left. The feature of such THGEM interface grid is that one has to apply a high enough voltage across it, to have the 100% electron transmittance. In particular in liquid Ar, for THGEM electrode with 28% optical transparency, the nominal electric field should exceed 5 kV/cm to provide the 100% electron transmission [79]: see Figure 21, right. On the other hand, it was shown there that the THGEM with 75% optical transparency provides 100% electron transmission at already 2 kV/cm of nominal electric field. The advantage of such THGEM electrodes as compared to wire electrodes, is that they are quite rigid which avoids the problem of wire grid sagging under high electric field in large-area two-phase detectors.

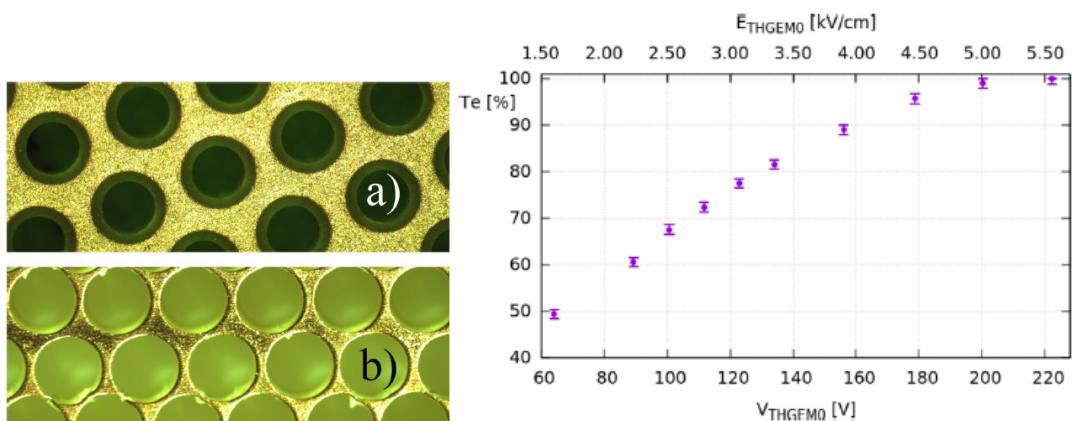

**Figure 21.** Left: microscope images of two THGEM electrodes used as interface grid in two-phase Ar detectors in [9,12]: with 28% and 75% optical transparency [79]. Right: Calculated electron transmission through 28% THGEM acting as an interface grid immersed in liquid Ar as a function of the voltage across THGEM (bottom axis) and nominal electric field in THGEM (top axis) [79]. The electric fields below and above the THGEM were 0.56 and 4.3 kV/cm respectively.

## 5. Combined charge/light signal amplification in the gas phase of two-phase detector, using avalanche scintillations

### 5.1. Two-phase Ar detector with combined THGEM/SiPM-matrix multiplier

Figure 22 shows the concept of combined charge/light signal amplification in two-phase detector with EL gap, using avalanche scintillations. In this concept, the combined THGEM/SiPM-matrix multiplier is coupled to the EL gap: the THGEM provides the charge signal amplification by operating in electron avalanching mode, while the SiPM matrix optically records avalanche scintillations produced in the THGEM holes thus providing the light signal amplification with position resolution. Since ordinary SiPMs produced by industry are typically sensitive to the NIR but not sensitive to the VUV, the avalanche scintillations can be recorded in two ways: either directly, using avalanche scintillations due to atomic transitions in the NIR (see Figures 4 and 5), or indirectly via WLS film in front of the SiPM matrix, using excimer avalanche scintillations in the VUV. The earlier works on realizing this concept in two-phase Ar [55] and Xe [56] detectors should be mentioned, albeit with smaller active area and SiPM matrix size.

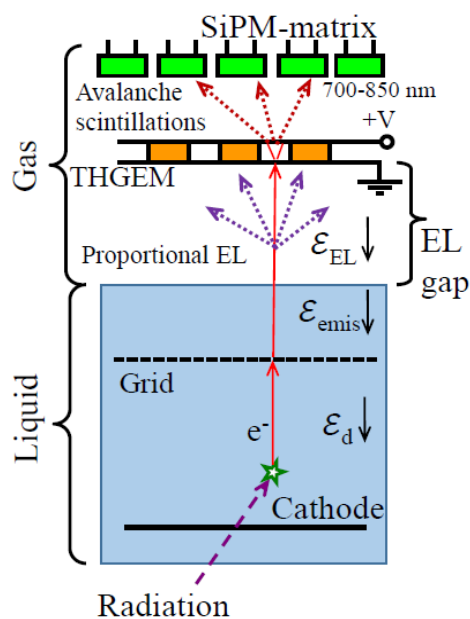

**Figure 22**. Concept of combined charge/light signal amplification in two-phase detector with EL gap, using avalanche scintillations and combined THGEM/SiPM-matrix multiplier [12].

In the most elaborated way, the concept was realized in two-phase Ar detector with combined THGEM/SiPM-matrix multiplier of 10x10 cm² active area in [12], using a 5x5 SiPM matrix having 1 cm pitch of SiPM elements. The amplitude and position resolution properties of the detector are illustrated in Figure 23. As large as about 500 photoelectrons were recorded by the SiPM matrix for 88 kev gamma-rays from ¹⁰⁹Cd source, at rather moderate THGEM charge gain, of 37. This would correspond to the detector yield of about 1-2 photoelectrons per drifting electron in the gas phase [12]. Moreover, the highest position resolution was obtained in this detector than ever was measured for two-phase detectors with EL gap: $\sigma = 26\text{mm}/\sqrt{N_{PE}}$, where $N_{PE}$ is the total number of photoelectrons recorded by the SiPM matrix. Such photoelectron yield and positions resolution would be sufficient for applications in neutrino and dark matter search experiments.

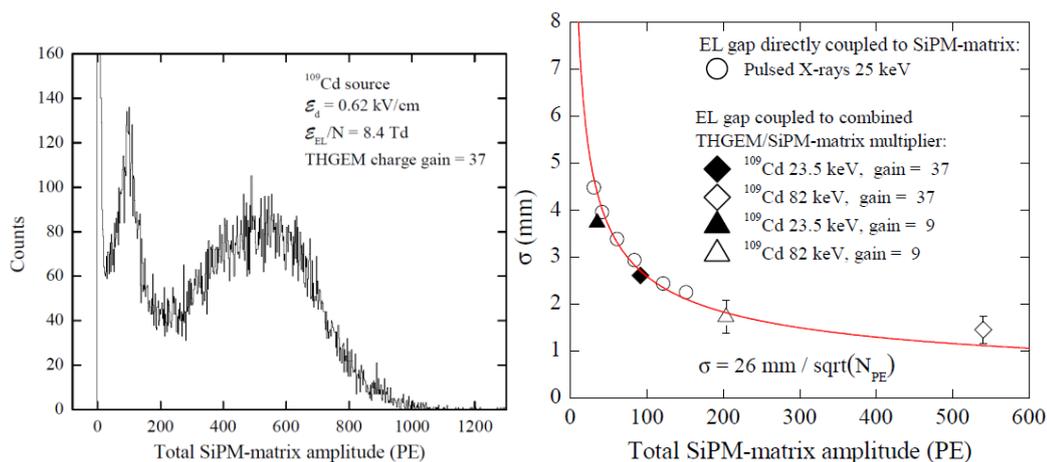

**Figure 23.** Total SiPM-matrix amplitude spectrum for gamma-rays of $^{109}$Cd source (left) and its position resolution (right), in two-phase Ar detector with EL gap and combined THGEM/SiPM-matrix multiplier, shown in Figure 22 [12]. The THGEM charge gain was 37. The two characteristic peaks of low (22-25 keV) and high energy (60-70 and 88 keV) lines of $^{109}$Cd source on W substrate are well separated in the amplitude spectrum. The position resolution (standard deviation) is shown as a function of the total number of photoelectrons recorded by the SiPM matrix. The curve is the fit by inverse root function.

### 5.2. Two-phase Ar detector with combined THGEM/CCD-camera multiplier

Figure 24 (left) shows the concept of combined charge/light signal amplification in two-phase Ar detector, using avalanche scintillations in combined THGEM/CCD-camera multiplier [60]. It is similar to that considered above, with the difference that the SiPM matrix is replaced by the CCD cameras. Here the avalanche scintillations were recorded indirectly via WLS film behind the THGEM plate, using excimer avalanche scintillations in the VUV. In the most elaborated way the concept was realized in [61] in two-phase Ar detector with 54x54 cm² active area and overall liquid Ar mass of 1 t. The advanced CCD cameras, with additional electronic amplification of the signal, namely Electron Multiplying Charge Coupled Device (EMCCD) were used, which allowed for single photoelectron counting.

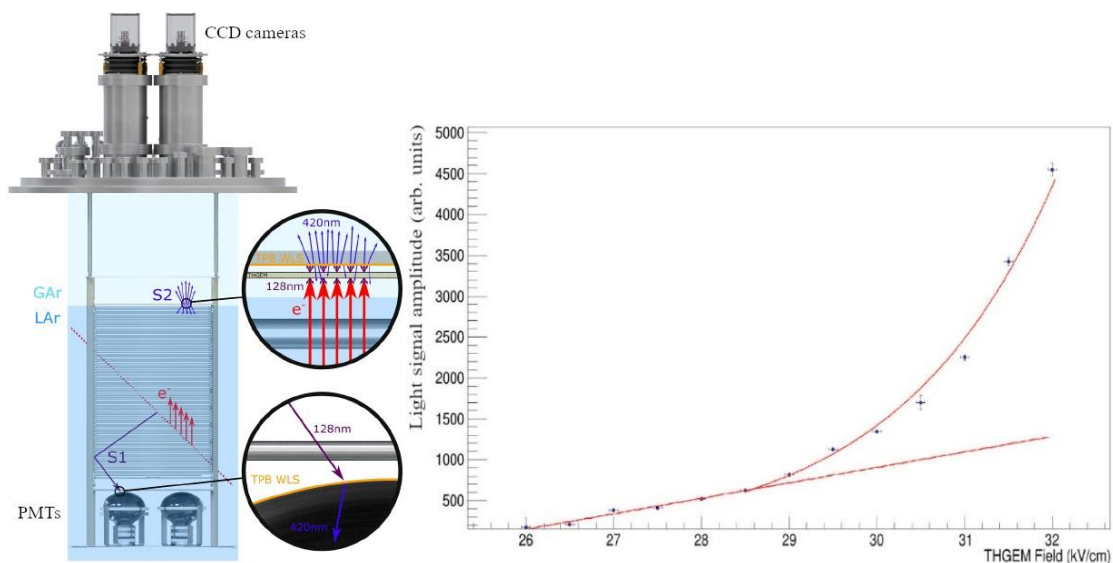

**Figure 24.** Left: concept of combined charge/light signal amplification in two-phase detector [61], using avalanche scintillations in combined THGEM/CCD-camera multiplier [60], in the most elaborated way realized in [61], with 54x54 cm² active area and overall liquid Ar mass of 1 t. Right: light signal amplitude on the CCD-cameras as a function of the nominal electric field in THGEM holes (data points) [61].

Using such detector, 3D reconstruction of cosmic muon tracks and beamline interactions has been successfully performed. The detector was proposed to be employed in future large scale two-phase liquid Ar neutrino experiments as an alternative readout method, replacing the highly segmented anode planes.

One can see from Figure 24 (right), showing the light signal amplitude dependence on the THGEM electric field, that two operation modes were observed: that of proportional EL (linear part) and that of electron avalanching (exponential rise part). The optical readout allows to measure the THGEM charge gain *ad hoc*: using the dependence of the light signal intensity on the electric field. In particular, at the maximum field of 32 kV/cm, the avalanche light gain, defined as the measured light intensity normalized to that of the linear extrapolation from the lower electric fields, is close to 4. This avalanche light gain corresponds to larger charge gain, because secondary (avalanche) electrons produce less EL light per drifting electron compared to initial electron, since they drift over smaller distance in the hole. If to approximate the THGEM hole by a parallel-plate gap, it can be shown that the light gain of 4 corresponds to the charge gain of 8.

## 6. Charge and light signal amplification in the liquid phase

### 6.1. Energy levels and scintillation mechanisms in liquid Ar

In the liquid, the excited, ground and ionized atomic states transform to the exciton, valence and conduction bands, respectively (see figure 3 and [7]), in particular in Ar, to the $Ar^*$ ($n = 1$, $^2P_{3/2}$) and $Ar^*$ ($n = 1$, $^2P_{1/2}$) excitons, which reflect the $^3P_1$ and $^1P_1$ levels of the $Ar^*(3p^54s^1)$ atomic states. According to reaction (15) of Table 1, the excitons are immediately trapped in singlet or triplet excimer states $Ar_2^*(^{1,3}\Sigma_u^+)$. The singlet $^1\Sigma_u^+$ and triplet $^3\Sigma_u^+$ excimers provide the fast and slow emission components in the VUV for scintillations in liquid Ar (reaction (16) of Table 1)) with a lifetime of 7 ns and 1.6 μs respectively (see [7] and references therein). This mechanism is the basic one for scintillations in liquid Ar and is similar to that of ordinary electroluminescence in gaseous Ar described in section 3.2. As for the other scintillation mechanism considered in that section, namely that of atomic transitions in the NIR in gaseous Ar, it does not work in liquid Ar due to the disappearance of analogues of the higher excited levels. On the other hand, there are some indications that the neutral bremsstrahlung emission may still exist in liquid Ar: see sections 6.2 and 6.3.

### 6.2. Charge and light signal amplification in liquid Xe and liquid Ar using wires, strips and needles

The attempts to obtain high and stable charge amplification directly in noble-gas liquids, using "open-geometry" multipliers, have not been very successful: rather low charge gains (≤10) were observed in liquid Xe using wire proportional counters [27,28,30] and micro-strip counter [29] and in liquid Ar using needle counter [31]. This is clear from Figure 25 showing charge gain in liquid Xe for different wire diameters as a function of the applied voltage.

On the other hand, proportional electroluminescence (EL) observed in liquid Xe using thin wires and VUV-sensitive PMTs [28,30] looks more promising for applications [62]: see Figure 26. In particular, the maximum proportional EL yield of about 300 photons per drifting electron was obtained using 10 μm diameter wire [30]. In a wide voltage range, proportional EL gain was well described by a linear function, defining a nominal EL threshold at a certain electric field. In particular, the nominal thresholds for proportional EL and electron multiplication were measured as about 400 and 700 kV/cm, respectively. The linear dependence of proportional EL gain on the electric field and the existence of the threshold for EL in liquid Xe were confirmed also in a needle-type device [80]: see Figure 27.

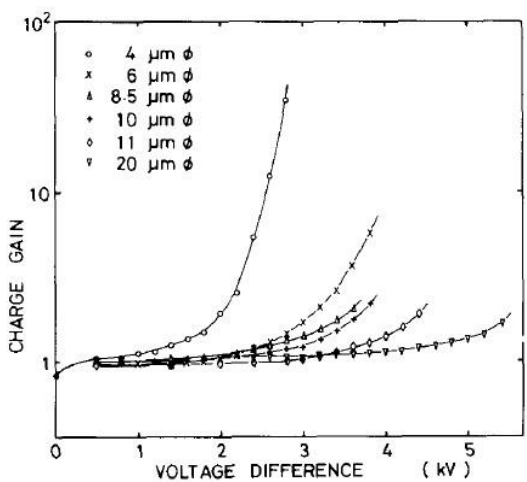

**Figure 25.** Charge gain as a function of applied voltage for different diameters of the anode wire in liquid Xe [28]. The maximum gains are limited by discharges.

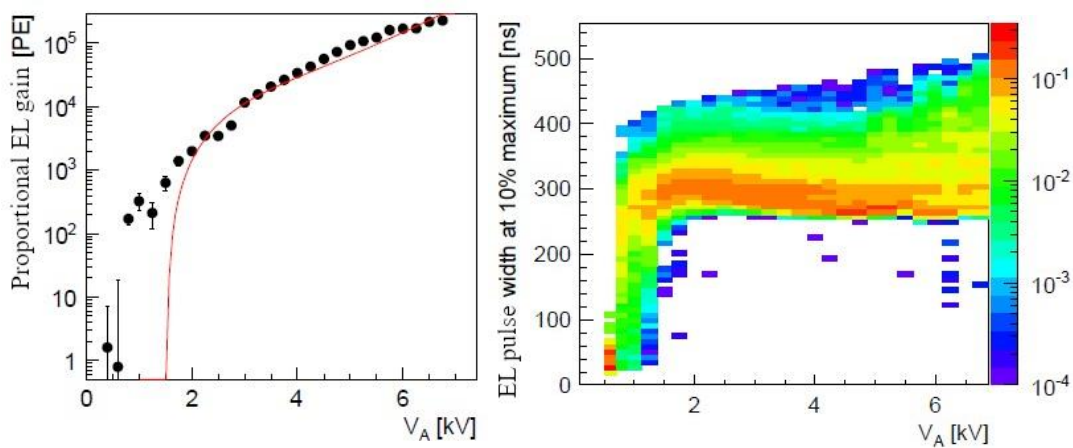

**Figure 26.** Left: proportional EL gain expressed in photoelectrons (PE) recorded by PMT as a function of applied voltage for 10 μm anode wire in liquid Xe [30]. The curve is the linear fit to the data points. Notice that the experimental data points diverge from this fit near and below the nominal EL threshold. Right: the EL pulse width at 10% of pulse maximum as a function of applied voltage [30]. Notice that the EL pulse width decreases below the nominal EL threshold, presumably indicating on alternative EL mechanism below the threshold.

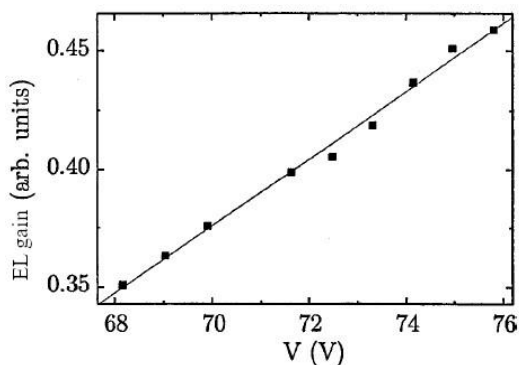

**Figure 27.** Proportional EL gain of needle-type device in liquid Xe as a function of the applied voltage [80]. The electric field near the tip of needle varies from $30 \times 10^3$ kV/cm to $2 \times 10^3$ kV/cm over distance of 0.1 μm. If to extrapolate to zero, light emission has a threshold at 44 V.

It is interesting however, that below the nominal EL threshold the EL signal was still observed in [30]: see Figure 26 (left). Moreover, the pulse width for such "alternative" EL dramatically dropped down compared to "ordinary" EL (see Figure 26 (right)), indicating that alternative EL is fast. This under-threshold and fast EL in liquid Xe looks very similar to under-threshold and fast EL in gaseous

Ar described in section 4.2, where it was convincingly explained by the neutral bremsstrahlung (NBrS) effect: compare Figure 26 (left) to Figures 5 and 9. This similarity has lead us to a conclusion that proportional EL in liquid Xe, observed under the nominal EL threshold, of about of 400 kV/cm, with high probability is due to the NBrS effect as well. Note that the similar hypothesis has been proposed in [9] to explain proportional EL in liquid Ar, observed in immersed THGEMs and GEMs at relatively low fields: see below.

### 6.3. Light signal amplification in liquid Ar using THGEMs and GEMs

In liquid Ar, proportional EL, with its characteristic properties of having a threshold and linear field dependence, was observed in immersed THGEM and GEM structures in [4,49]: see Figure 28. The mystery of the results obtained is that liquid Ar EL started at much lower electric fields compared to liquid Xe, at about 60 kV/cm, which almost an order of magnitude lower than that of liquid Xe. In addition, these electric fields are far less than those of $3 \times 10^3$ kV/cm expected from theoretical calculations for EL in liquid Ar due to atomic excitation mechanisms [81].

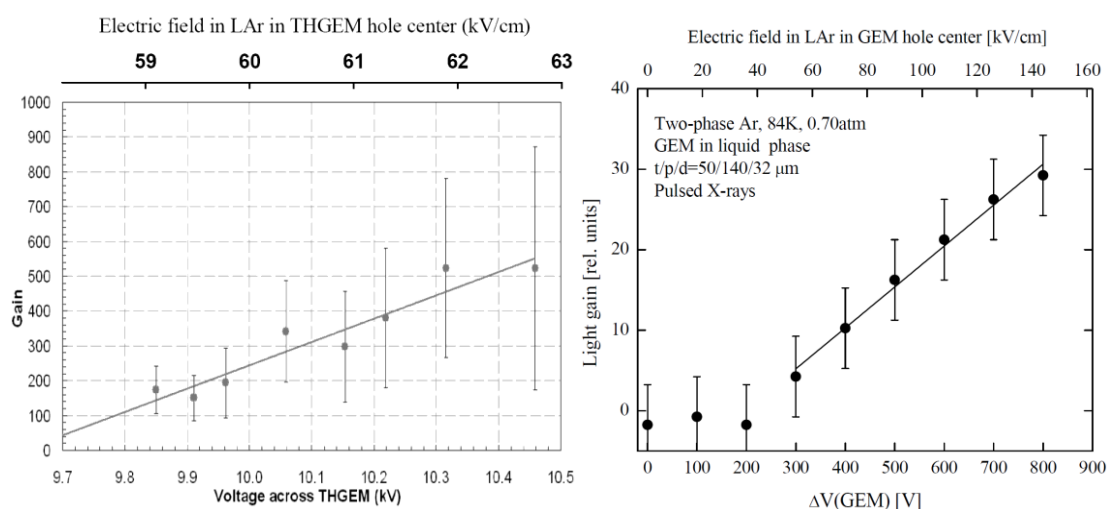

**Figure 28.** Relative amplitude of the proportional EL signal generated in THGEM (left) [49] and GEM (right) [4] immersed in liquid Ar as a function of the applied voltage [4]. The appropriate electric fields in the THGEM and GEM hole centers are shown on the top axes.

It was supposed in [9] that the NBrS effect could be responsible for proportional EL in liquid Ar observed in immersed GEM-like structures. Indeed, the electric fields in the center of GEM or THGEM holes used in liquid Ar, of 60–140 kV/cm, correspond to E/N=0.3-0.7 Td that are not that small. For such reduced electric fields, the theory predicts that NBrS EL already exists: see Figure 9. It also predicts the linear dependence of the EL yield observed in experiment. Thus, one cannot exclude the NBrS origin of proportional EL in noble-gas liquids at lower electric fields.

Another possible explanation might the presence for some reason of gaseous bubbles associated to the THGEM and GEM holes, within which proportional EL in the gas phase could develop. This effect will be discussed below. This hypothesis, of bubble-assisted EL, could be confidently tested by measuring the EL emission spectrum in the NIR: it would be confirmed if the emission spectrum would consist of atomic lines corresponding to gaseous Ar. Otherwise, the continuous spectrum would indicate on the real liquid Ar emission. This method has been realized in [82] when studying electrical breakdown in liquid Ar; the results will be discussed in section 6.5.

### 6.4. Liquid-Hole Multipliers in liquid Ar and liquid Xe

The Liquid-Hole Multiplier (LHM) concept [57,58] has emerged after observation of proportional EL in THGEMs immersed in liquid Xe: see Figure 29 [83]. Notice the linear voltage dependence of EL yield and rather low voltage threshold for EL, corresponding to the electric field in THGEM holes of about 10 kV/cm. This field value is one and two orders of magnitude smaller than that for proportional EL in liquid Ar using THGEMs and in liquid Xe using wires, respectively. Therefore such a low threshold EL can hardly be explained by ordinary EL or NBrS EL. This effect was proved to be due to proportional EL in the bubbles formed in the holes of the THGEM and under it [58,84].

In the LHM concept, shown in Figure 30, the noble-gas bubble is formed in noble-gas liquid under the THGEM or GEM plate in a controlled way using heating wires. Radiation-induced ionization electrons and scintillation (S1)-induced photoelectrons from a CsI photocathode (optionally deposited on top of the THGEM) are focused into the THGHEM holes; both cross the liquid-vapor interface and create EL signals (S2 and S1′, respectively) in the bubble.

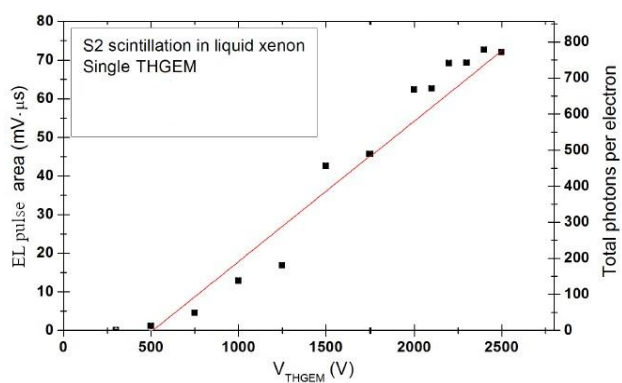

**Figure 29.** Amplitude of proportional EL signal (pulse-area) and its EL yield (number of photons per drifting electron) produced by THGEM in liquid Xe as a function of the THGEM voltage [83].

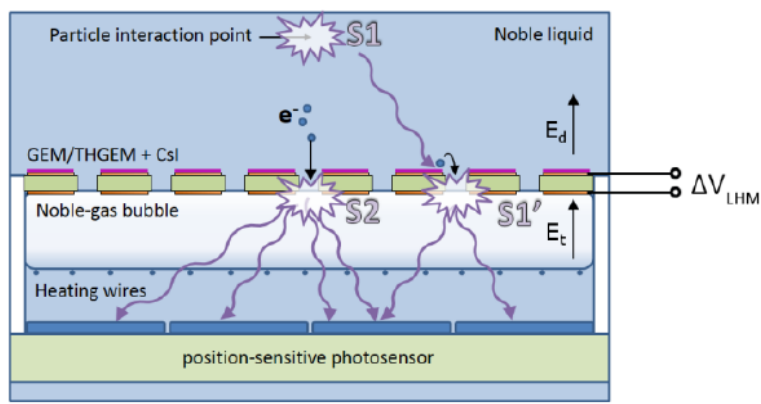

**Figure 30.** Concept of LHM detector [57,58,59].

In some sense, the LHM bubble resembles the DarkSide scheme [85], where a gas pocket is formed using heating wires, confined from above by a metallized quartz-plate anode. However, there is a fundamental difference: the gas bubbles in LHM are kept from coming up through the THGEM holes due to surface tension. This allows ionization to enter the gas phase from both sides of the liquid: from below and above the bubble.

Figure 31 illustrates the LHM performance in liquid Xe and liquid Ar: the EL (S2) signals are distinctly seen along with the primary scintillation (S1) signals. In liquid Xe, the EL yield of as high as 400 photons per drifting electron was reached, which is comparable with that obtained in traditional two-phase Xe detector with 1 cm thick EL gap. Nevertheless, further research is needed before this technique can be applied in real rare-event experiments.

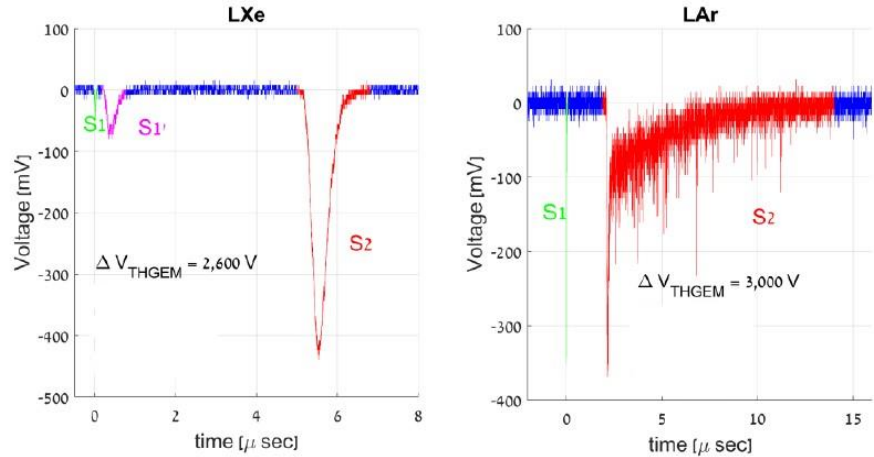

**Figure 31.** Example of EL signals (S2 signals) induced by alpha particles, recorded from a LHM detector [59]: (left) in liquid Xe and bare-PMT and (right) in liquid Ar and TPB-coated PMT.

*6.5. Breakdowns in noble-gas liquids*

Breakdowns in noble-gas liquids are relevant to the subject of this review because they are accompanied by both charge multiplication and electroluminescence. Their study is important for practical implementation of large liquid Ar TPCs for neutrino experiments, where the high electrical potentials are envisioned, from 100 kV to 1 MV.

The breakdown mechanism is not fully understood; it includes a number of physical effects like electron field emission, streamer propagation and electroluminescence. Nevertheless, the universal empirical rule was revealed in breakdown measurements at cm scale [82,86,87,88]: the threshold for dielectric breakdown in any noble-gas liquid decreases with the increase of the surface area of the electrodes according to inverse power law:

$$E_{max} = C \cdot A^p \quad , \tag{11}$$

where $C$ is a material-dependent constant, $A$ is the stressed area (defined as the area with an electric field intensity above 90% of its maximum), and $p \approx -0.25$ . See also Figure 32 showing the breakdown field in liquid Ar as a function of stressed area. There does not appear to be a significant difference between breakdown behavior in liquid Ar and liquid Xe [88].

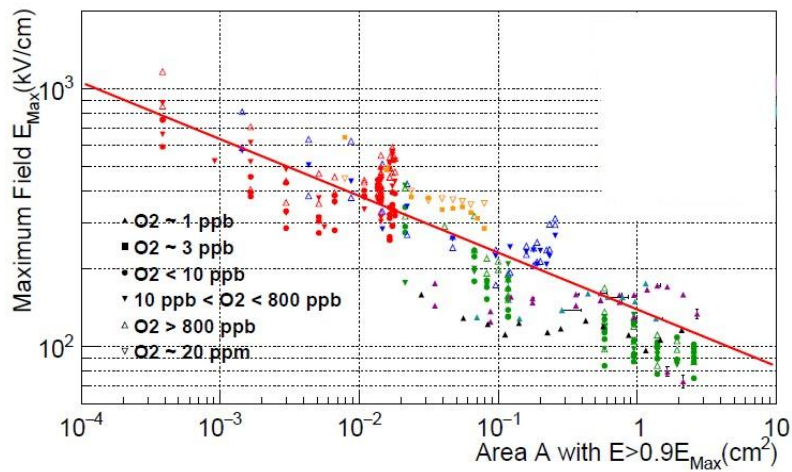

**Figure 32.** Breakdown field versus stressed area of the cathode in liquid Ar [82]. The stressed area is defined as the area with electric field greater than 90% of the maximum electric field in the gap. The fit line represents the dependence $E_{max} = C \cdot A^p$ with $C = 139$ and $p = -0.22$.

It is remarkable that the breakdown in liquid Ar was proposed to have a two-phase nature [82]: while at its first stage electron electroluminescence and avalanching occurred directly in the liquid (induced by field emission of the electrons from the cathode), at the second stage the streamer propagated in the elongated gas bubble. This hypothesis is supported by Figure 33. It shows photon emission spectra in the visible range for a breakdown in liquid Ar at the first (field emission) and the second (streamer) stage. The field emission stage was observed in the form of a glowing cone. These spectra feature the presence of both a liquid and a gas phase in the breakdown. Indeed, the curve with a broad continuum resembles the scintillation continuous spectrum observed in the visible range in liquid Ar [89], while the curve featuring distinct peaks around 700 and 750 nm is attributed to the line spectrum in the NIR due to $Ar^*(3p^54p^1) \rightarrow Ar^*(3p^54s^1)$ transitions in gaseous Ar (see Figure 4 and Equation (9)).

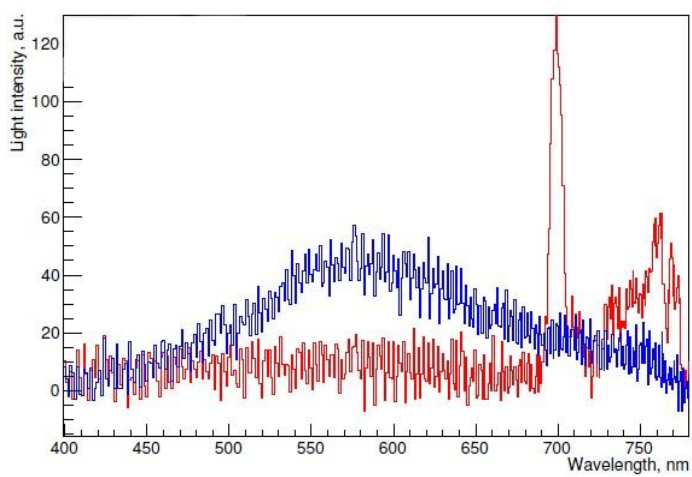

**Figure 33.** Photon emission spectra of electroluminescence at the first (field emission cone) stage and the second (streamer) stage for a breakdown in liquid Ar [82]: the curves are that of blue with a broad continuum and that of red with distinct peaks, respectively. The first stage curve, with a broad continuum, is similar to the scintillation continuous spectrum of liquid Ar observed in [89], while the second stage curve, featuring distinct peaks around 700 and 750 nm, is attributed to the line spectrum of $Ar^*(3p^54p^1) \rightarrow Ar^*(3p^54s^1)$ transitions in gaseous Ar.

## 7. Conclusions

In this work we have reviewed physics and techniques of direct signal amplification in the detection media of two-phase argon and xenon detectors. Originally given rise from the idea of the two-phase detector itself, the concept of amplifying the primary ionization signal (S2 signal) had triggered intense and difficult R&D work in the course of the last 20 years. This resulted in a large variety of advanced concepts of S2 signal amplification developed in this period. All these concepts are based on two physical effects: electroluminescence and electron avalanching under high electric fields, both in the gas and in the liquid phase.

For the time being the most intensively used concept (in particular in dark matter search experiments) is that of the "light signal amplification" utilizing proportional electroluminescence in the gas phase. Another most intensively studied concept is that of the "charge signal amplification" using electron avalanching in THGEM multipliers. This however encounters difficulties with large active areas needed for large-scale liquid Ar detectors. The solution of the problem might be to use combined charge/light signal amplification.

In addition, this R&D work has significantly advanced understanding of physical effects related to amplification processes in two-phase detectors. Among them are new electroluminescence mechanisms, including that due to neutral bremsstrahlung effect.

Finally, such kinds of techniques may come to be in great demand in rare-event experiments, such as those of dark matter search, coherent neutrino-nucleus scattering and giant liquid Ar detectors for neutrino physics. Further studies in this field are in progress.

**Funding:** The minor part of the work (section 4), was supported by Russian Foundation for Basic Research (project no. 18-02-00117). The major part of the work (sections 1-3 and 5-7) was supported by Russian Science Foundation (project no. 20-12-00008).

**Acknowledgments:** This review was triggered by author's participation in preparing a forthcoming book "Two-phase emission detectors", in collaboration with D. Akimov, A. Bolozdynya and V. Chepel, to be published by World Scientific Publishing Co. Pte. Ltd.